\documentclass[conference]{IEEEtran}
\usepackage[nospace]{cite}
\usepackage{xspace}
\usepackage{graphicx}
\usepackage{subfig}
\usepackage{blindtext, xcolor}
\usepackage{comment}
\graphicspath{ {./figures/} }
\usepackage[justification=centering]{caption}
\usepackage[]{algorithmic}
\usepackage{caption}
\usepackage[ruled,vlined]{algorithm2e}
\usepackage{amssymb}
\usepackage{amsmath}

\newcommand{\CASE}[1]{\STATE \textbf{case} #1\textbf{:} \begin{ALC@g}}
\newcommand{\ENDCASE}{\end{ALC@g}}

\newcommand{\DEFAULT}{\STATE \textbf{default:} \begin{ALC@g}}
\newcommand{\ENDDEFAULT}{\end{ALC@g}}
\newcommand{\DEFAULTLINE}[1]{\STATE \textbf{default:} }
\newcommand{\elem}[1]{\textsf{\small{#1}}}
\newcommand{\mRUBiS}{\mbox{mRUBiS}\xspace}
\newcommand{\scalefactor}{0.99}
\newcommand{\Pa}[0]{\mathcal{P}}

\usepackage{url}
\makeatletter
\def\ps@IEEEtitlepagestyle{
	\def\@oddfoot{\mycopyrightnotice \thepage}
	\def\@evenfoot{}
}
\def\mycopyrightnotice{
	{\footnotesize
		\begin{minipage}{\textwidth}
			\centering
			\textcopyright~2016 IEEE. Personal use of this material is permitted.
			Permission from IEEE must be obtained for all other uses, in any current or future media, including reprinting/republishing this material for advertising or promotional purposes, creating new collective works, for resale or redistribution to servers or lists, or reuse of any copyrighted component of this work in other works. 
			{\tt DOI:} \url{https://doi.org/10.1109/SASO.2016.21}
		\end{minipage}
	}
}
\usepackage[hidelinks]{hyperref}
\hypersetup{
	bookmarks=true,         
	unicode=false,          
	pdftoolbar=true,        
	pdfmenubar=true,        
	pdffitwindow=false,     
	pdftitle={Towards Linking Adaptation Rules to the Utility Function for Dynamic Architectures},
	pdfauthor={Sona Ghahremani, Holger Giese and Thomas Vogel},
	pdfsubject={},
	pdfcreator={},
	pdfproducer={},
	pdfkeywords={self-healing, adaptation rules, architecture-based adaptation, utility},
	pdfnewwindow=true,
}

\begin{document}

\title{Towards Linking Adaptation Rules to the Utility Function for Dynamic Architectures}

\author{
\IEEEauthorblockN{Sona Ghahremani, Holger Giese and Thomas Vogel}
\IEEEauthorblockA{Hasso Plattner Institute, University of Potsdam, 
Prof.-Dr.-Helmert-Str. 2-3, D-14482 Potsdam, Germany\\
Email: \{sona.ghahremani$|$holger.giese$|$thomas.vogel\}@hpi.uni-potsdam.de}
}

\maketitle
\pagestyle{plain}

\IEEEpeerreviewmaketitle
\begin{abstract}
To benefit from utility-driven and rule-based approaches to self-adaptation, we propose combining both by defining and linking the utility function and the adaptation rules in a pattern-based way at the architectural level.
\end{abstract}
\section{Introduction}\label{sec:intro}
\noindent
To realize self-adaptation, a software system is equipped with a feedback loop that \underline{m}onitors and \underline{a}nalyzes the system and if needed, \underline{p}lans and \underline{e}xecutes adaptation to the system. For this purpose, the feedback loop uses \underline{k}nowledge (cf. \textit{\mbox{MAPE-K}}~\cite{Kephart&Chess2003}).
To achieve architectural self-adaptation, the feedback loop maintains a \textit{runtime model}~\cite{Blair+2009}, as part of its knowledge, which represents the architecture of the system under adaptation.
This paves the way for dynamic architectures, in which issues can be identified and handled by architectural self-adaptation~\cite{2012ChengStitch}. 

There are various ways of how self-adaptation can be realized. 
On the one hand, \emph{rule-based} approaches~\cite{1537890,Keeney:2003:CPC:826036.826854} prescribe the adaptation (i.e., actions) to be executed if specific events occur and if specific conditions are satisfied. Therefore, the adaptation rules are encoded as event-condition-action rules. Such approaches employ the predefined adaptation rules of a rule set $\Re$ and usually result in a scalable solution, however, often only with at most sufficient adaptation decisions. 
On the other hand, \emph{utility-driven} approaches~\cite{Kephart+Walsh2004,Esfahani+2013} often enable optimal decisions by searching the possible space of adaptations to find the optimal one according to a \emph{utility function} $U$, but usually do not scale well for larger problems. A utility function $U$ is an objective policy which expresses the value of each possible configuration of the system in its domain and identifies the degree to which \emph{goals} of the system have been satisfied.

Extensive research has been made on utility functions and utility-driven decision-making policies that operates at the level of the software architecture and architectural features. 
However, the utility-driven approaches proposed in literature all pursue a search-based optimization in the solution space that often does not scale well for complex systems with large solution spaces~\cite{Esfahani+2013,RouvoyEtAl09}, or they define the utility function over a pre-defined and hence bounded configuration space~\cite{2012ChengStitch,1128711}.

Consequently, we propose combining rule-based and utility-driven approaches to achieve the beneficial properties of each of them for dynamic architectures. 
Defining the utility function and the adaptation rules in a pattern-based way at the architectural level allows us to combine both approaches and particularly to estimate the impact of each application of an adaptation rule on the overall utility at runtime. 

The paper is organized as follows:
In Sec.~\ref{sec:utility}, we introduce the pattern-based definition of utility functions for dynamic architectures.
In Sec.~\ref{subsec:link}, we link the adaptation rules to the utility function and discuss related observations we exploited for the implementation.
Finally, we report on a preliminary evaluation, conclude, and outline future work.

\section{Utility Functions for Dynamic Architectures}\label{sec:utility}
\noindent
Desirable or undesirable issues for a dynamic architecture can be expressed as positive respectively negative model \emph{patterns} such that concrete issues relate to occurrences of these patterns in a runtime model $G$. We denote that an occurrence as a match $m$ for a pattern $P$ in the runtime model $G$ exists by $G \models_m P$.
Therefore, we propose defining a utility function for a dynamic architecture represented in a runtime model with patterns.

For any utility function for a dynamic architecture must hold that (i) the optimal architectural configuration where all the system goals are optimally fulfilled must gain the maximum utility and that (ii) if any constraint or goal is violated, it must lead to a decrease of the utility. 
Thus, we employ \emph{positive architectural utility patterns} $\Pa^+ = \{ P^+_1, \dots, P^+_k \}$ and capture their impact on the utility accordingly using utility sub-functions $U^+_i$ to address case (i). 
Similarly, we employ \emph{negative architectural utility patterns} $\Pa^- = \{ P^-_1, \dots, P^-_n \}$ and capture their impact on the utility accordingly using utility sub-functions $U^-_j$ to address case (ii). 
It has to be noted that the impact of each pattern occurrence on the overall utility, which is captured by a utility sub-function, may vary for each occurrence depending on the specific characteristics of the context (e.g., for self-healing, the attributes of a failing component may indicate the criticality of this component or the severity of the observed failures, which determines the negative impact on the utility). 

As an example, we use \textit{\mRUBiS}~\cite{mRUBiS}, a component-based system realizing an online marketplace that hosts an arbitrary number of shops. Each shop consists of 18 components and runs isolated from the other shops. 
We are particularly interested into self-healing, that is, the automatic repairing of runtime failures by architectural self-adaptation. 
Therefore, we equip \mRUBiS with a \mbox{MAPE-K} loop that uses a runtime model representing the \mRUBiS architecture. 
The model captures the hosted \elem{Shop}s, their structuring into \elem{Component}s (with \elem{ProvidedInterface}s and \elem{RequiredInterface}s) and \elem{Connector}s linking required and provided interfaces, as well as runtime \elem{Failure}s. Thereby, each \elem{Component} is of a specific \elem{ComponentType}.

\begin{figure}[h]
\vspace{-1mm}
\begin{centering}
 \includegraphics[scale=\scalefactor]{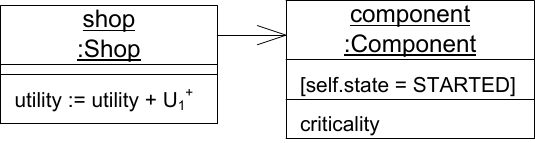}
 \vspace{-1mm}
  \caption{Positive Architectural Utility Pattern $P^+_1.$ }
  \label{fig:Pospatt}
  \end{centering}
\vspace{-4mm}
\end{figure}

For the example, Fig.~\ref{fig:Pospatt} shows the positive architectural utility pattern $P^+_1$ with its utility sub-function $U^+_1$. For each started (i.e., running) component of a shop, the utility of the shop is increased by $U^+_1$. We may define $U^+_1$ as the product of the criticality of the specific component, the reliability of the component type, and the connectivity of the component (i.e., the number of associated connectors). 
This pattern is applied to all shops on the marketplace to obtain the total positive impact on the overall utility of the marketplace.

\begin{figure}[h]
\vspace{-1mm}
\begin{centering}
  \includegraphics[scale=\scalefactor]{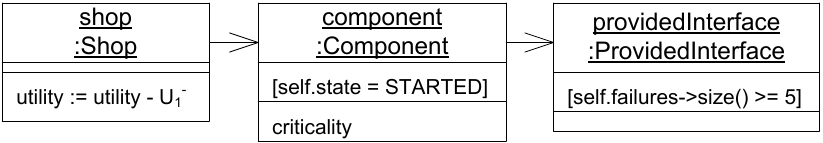}
  \vspace{-1mm}
  \caption{Negative Architectural Utility Pattern $P^-_1$.}
  \label{fig:Antipatt}
  \end{centering}
\vspace{-4mm}
\end{figure}

Similarly, Fig.~\ref{fig:Antipatt} presents the negative architectural utility pattern $P^-_1$ with its utility sub-function $U^-_1$ for the example. This pattern matches if the usage of a provided interface of a started component in a shop has caused five or more failures (exceptions), which decreases the utility of the shop by $U^-_1$. We may define $U^-_1$ similar to $U^+_1$ and apply this pattern for all shops to obtain the total negative impact on the overall utility.

In general, we define \textit{multiple} positive and negative patterns, that is, $\Pa^+ = \{ P^+_1, \dots, P^+_k \}$ and $\Pa^- = \{ P^-_1, \dots, P^-_n \}$.
All matches of these patterns determine the overall utility $U(G)$ for the current architecture represented in the runtime model $G$. 
We define the set of all matches for the positive pattern $P^+_i$ in $G$ as $M^+_i(G) = \{ m | P^+_i \models_m G \}$ and the set of all matches for the negative pattern $P^-_j$ in $G$ as $M^-_j(G) = \{ m | P^-_j \models_m G \}$. Given these sets, the overall utility $U(G)$ can be defined and computed as follows:
\vspace{-.5mm}
\begin{equation}\label{eq:utility}
  U(G) 
:=
  \sum_{i=1}^{k}
      \sum_{m \in\!M^+_i\!(G)}
        \!\!\!\!\!U^+_i\!(G,m_i)
 -
  \sum_{j=1}^{n}
      \sum_{m \in\!M^-_j\!(G)}
        \!\!\!\!\!U^-_j\!(G,m_j)
\end{equation}
Hence, the overall utility is the sum of all $U^+_i$ for each match in $M^+_i(G)$ accumulated over all $k$ positive patterns in $\Pa^+$ 
minus the 
sum of all $U^-_j$ for each match in $M^-_j(G)$ accumulated over all $n$ negative patterns in $\Pa^-$.
As discussed previously, the impact of a match on the overall utility is influenced by the specific context of the match. Thus, the utility sub-functions $U^+_i$ and $U^-_j$ are paramterized by the runtime model $G$ and the specific match $m$, which capture the context of the match.

In general, matches of positive and negative patterns result from changes of the environment.
While existing matches of positive patterns are usually not affected by the adaptation rules (i.e., the adaptation does not do any harm to the system), matches of negative patterns should be addressed by the rules (i.e., the adaptation repairs the system).
Finally, the kind of utility functions as presented here is restricted to linear functions, which are often used for optimization (cf.~\cite{1128711}).

\section{Linking Adaptation Rules\,\&\,Utility Functions}\label{subsec:link}

\noindent
Based on the idea of architectural utility patterns, we describe the condition of an adaptation rule $r$, which must be fulfilled in order to apply $r$, as a model pattern $P$.
For a rule $r=(P,\dots)$ with the pattern $P$, we denote that $r$ can be applied for a match $m$ of $P$ in the runtime model $G$ as $G \models_m P$. Actually applying $r$ for $m$ in $G$ results in a modified runtime model $G'$, which we denote as $G \rightarrow_{r,m} G'$.

For any adaptation logic based on the presented kind of utility functions and adaptation rules, the following observations must hold:
(1) If there is no match of any negative pattern, there is no need for any adaptation since no improvement of the utility is possible. 
(2) Any possible improvement of the utility must necessarily resolve matches of negative patterns, otherwise no improvement of the utility would be possible. 

Consequently, we can assume that (A1) for any rule $r_o = (P_o, \dots)$ in the rule set $\Re$ must hold that a negative pattern $P^-_j$ exists such that any match $m_o$ for $P_o$ includes a match $m_j$ for $P^-_j$. Otherwise, $r_o$ could be applied even though no utility improvement can be achieved. This would contradict observation (1). 
However, $P_o$ might require more context compared to $P^-_j$ such that patterns of rules may extend the negative patterns.
Furthermore, it is plausible to assume that (A2) for any rule $r_o = (P_o, \dots)$ in $\Re$ and any match $m_o$ for $P_o$ that includes a match $m_j$ for the related negative pattern $P^-_j$ holds that applying $r_o$ for $m_o$ will make the match $m_j$ for $P^-_j$ invalid. Otherwise, $r_o$ would not resolve the identified match for $P^-_j$ and thus would not lead to an improvement of the utility as we would expect from observation (2).

\section{Discussion, Conclusion, \& Future Work}\label{sec:discussion}

\noindent
For a preliminary evaluation of the approach, we set up \mRUBiS with $100$ shops resulting in a system with $1800$ components. For this large case study, we were able to follow the proposed scheme by defining accordingly a utility function and linking several self-healing rules (e.g., restart, redeploy, replace a component) to the function. This gives us initial confidence about the applicability and benefits of the approach. 

In this paper, we proposed combining rule-based and utility-driven approaches to achieve the beneficial properties of each of them. Therefore, we define both, the utility function and the adaptation rules in a pattern-based way at the architectural level, which enables their linking.
As future work, we plan to exploit this link for the efficient and optimal execution of the rules, and to investigate non-linear utility functions.


\end{document}